# Universal shapes of self-organized semiconductor quantum dots

G. Costantini[1], A. Rastelli[1], C. Manzano[1], R. Songmuang[1], O.G. Schmidt[1], H. v. Känel[2], and K. Kern[1]

[1]*Max-Planck-Institut für Festkörperforschung, Heisenbergstr.1, D-70569 Stuttgart, Germany*

[2]*INFM and L-NESS, Dipartimento di Fisica, Politecnico di Milano a Como, Via Anzani 2, I-22100 Como, Italy*

The model systems for self-organized quantum dots formed from elemental and compound semiconductors, namely Ge grown on Si(001) and InAs on GaAs(001), are comparatively studied by scanning tunneling microscopy. It is shown that in both material combinations only two well-defined families of faceted and defect-free nanocrystals exist (and coexist). These three-dimensional islands, pyramids and domes, show common morphological characteristics, independently of the specific material system. Universal behavior is also observed in the capping-passivation process that turns the nanocrystals in true quantum dots.

To whom correspondence should be addressed. E-mail: gio@fkf.mpg.de

Self-organized semiconductor three-dimensional (3D) islands, epitaxially grown on lattice-mismatched substrates, are promising candidates for the practical realization of "artificial atoms", or so-called quantum dots (QDs). Their peculiar tunable properties open the way to novel applications in the fields of optoelectronics (*1*), single-electron (*2*) and single-photon (*3*) devices as well as quantum computation (*4*). However, a successful implementation requires a precise control over the density and the position of the islands and a good uniformity of their shapes and sizes. While the first two subjects have been successfully addressed (*5-7*), the control of shape and size is still an open problem. It entails a basic understanding of the actual morphology of the islands and of their further evolution during post-growth treatments, which for most material systems are still not fully settled issues.

The field of self-organized semiconductor quantum dots is dominated by two model systems, Ge/Si(001) and InAs/GaAs(001), where most of the devised applications have been developed and tested. Both have been thoroughly investigated, but only for the former a coherent picture of the spontaneously formed 3D islands that act as quantum dots has emerged. Shallow islands bound by {105} facets and steeper dome-like structures (*8*) are the only faceted and dislocation-free islands forming over the entire range of deposition parameters. On the contrary, the picture is much more confused for the InAs/GaAs(001) system, where a large variety of sizes and shapes has been reported (*1, 9-14*). Here we demonstrate that the existence (and coexistence) of only two families of faceted and defect-free islands (pyramids and domes) as well as the transformations that these undergo during the embedding-passivation process, are universal features and do not depend on the specific material system.

We used magnetron sputtering epitaxy for the growth of Ge on Si(001) and molecular beam epitaxy for that of InAs on GaAs(001). The substrates were first de-oxidized and then overgrown with a thick buffer layer of Si and GaAs, respectively. For the former material system, the QD samples were prepared by depositing 7.0 monolayers (ML) of Ge at a temperature of 550°C and a deposition rate of 0.3 ML/s. For the latter, 1.8 ML of InAs were deposited at 500°C, at a rate of 0.008 ML/s and with an As beam equivalent pressure of $8 \cdot 10^{-6}$ mbar. Immediately after stopping the Ge (In) flux, the samples were cooled to room temperature, transferred under ultra high vacuum conditions to a scanning tunneling microscope (STM), and there analyzed at room temperature.

Both material systems follow the Stranski-Krastanow growth mode, i.e. the 3D islands form only after the completion of a flat wetting layer with a thickness of about 3 and 1.7 ML for Ge/Si(001) and InAs/GaAs(001), respectively. The chosen deposition conditions are close to the thermodynamic limit (*15*) (high temperatures and low deposition rates (*16*)) and result in the coexistence of small and large islands (Figs. 1A and 1D). In analogy to the nomenclature used for Ge/Si(001), we shall generally call them *pyramids* and *domes*. The grayscale in Figs. 1A and 1D corresponds to the local surface slope computed as $\tan^{-1}(|\mathbf{n}|)$, where $\mathbf{n} = \nabla f$ is the surface gradient and f(x,y) is the surface height at position (x,y) (*17, 18*). A first visual inspection already reveals that in both systems domes are composed by different steep facets (darker regions) while pyramids are bound only by shallow facets (lighter regions). In order to obtain quantitative values, we plot in a two-dimensional histogram the frequency at which the values of $\mathbf{n}$ appear in STM images. In this way, all the points associated with a given surface orientation contribute to the same spot in the resulting intensity plot, also referred to as facet plot (FP) (*17, 18*). The application of this analysis to the Ge/Si(001) pyramid islands, produces four symmetrically arranged spots (Fig. 1B), whose distance and angular position from the central (0,0) point reveal a {105} orientation. On the other hand, domes produce 16 spots corresponding to {105}, {113} and {15 3 23} facets (*18*) (see Fig. 1C). The InAs/GaAs(001) data show a striking similarity both in the STM topographies (Fig. 1D) and in the corresponding FPs (Figs. 1E and 1F). The same type of structures can be recognized, with the difference that the shallow facets of pyramid islands are now oriented along the four non-perpendicular {137} orientations, while the domes produce 12 spots corresponding to {137}, {101} and {111} facets. We incidentally note that the FP technique allows a high-precision determination of the facet orientation which, in the case of Fig. 1E, explicitly excludes the similar {113}, {215} and {136} orientations, previously proposed for shallow InAs QDs (*12-14*).

The detailed morphological description of the QDs shown in Fig. 2 (directly derived from the facet analysis in Fig. 1), represents an important clarification for the InAs/GaAs(001) system. In fact, although being the most studied system with the most promising applications, before this work a simple coherent picture was missing. The model agrees with several reports on bimodal size distributions and with previous sparse facet assignment based on low-precision aspect ratio measurements (*10*) or diffraction experiments (*1, 11, 14*).

Furthermore, the results in Fig. 1 are the experimental proof that two universal island shapes - pyramid and dome – exist, as theoretically predicted by Daruka et al. (*19, 20*). Based on the minimization of surface and volume strain energy, these authors found out that islands in thermodynamic equilibrium should undergo a first order shape transition at a critical volume with the introduction of steep facets. This picture nicely agrees with the experimental observations reported here. Even the presence of one type of shallow facets at the base and at the apex of the domes with the same orientation of those bounding the pyramids, is actually observed in both material systems (see Figs. 1 and 2). Based on this theory, it becomes possible to extend the similarities between Ge/Si(001) and InAs/GaAs(001) QDs and to advance some predictions for the latter material system. As for the Ge/Si system (*8, 21*), also in the case of InAs/GaAs a transition between pyramids and domes should take place around a certain critical volume by increasing the amount of deposited material or upon annealing. The results reported in Ref. (*10*) and Ref. (*11*) respectively, can actually be reinterpreted in this light. Moreover, under appropriate growth conditions, transition islands (*21, 22*) with an intermediate shape between pyramids and domes should be observed. Further, the deposition of an $In_xGa_{1-x}As$ alloy should produce the same type of islands shown here, even if these are expected to have a larger size because of the inverse dependence of the transition critical volume on the strain energy (*23, 24*). Finally, on the basis of the results in Fig. 1, also the monodisperse population of Ge pyramids reported for small amounts of deposited material and/or low growth temperatures (*25, 26*), appears as the natural counterpart of the monomodal {137} InAs pyramid distribution measured by Márquez et al. (*9*).

The remarkable similarity between the Ge/Si(001) and InAs/GaAs(001) systems extends also to the capping of self-organized 3D islands. This procedure, mandatory to complete the quantum dot structures for device applications and for protecting them against degradation, typically induces strong structural modifications of the pristine islands, which finally determine the real electronic properties of the quantum dot. A detailed description of this phenomenon is therefore highly desirable, but is at present only available for the Ge/Si system (*24*). In this work the overgrowth experiments were done by depositing different amounts of Si (GaAs) at a temperature of 450°C (460°C) and a deposition rate of 0.1 ML/s (0.6 ML/s) on Ge (InAs) islands as those shown in Figs. 1A and 1D. For both systems the relevant morphological transformations occur during the deposition of the very first MLs and produce a very rapid decay of the island height (*24, 27, 28*). Thereafter an almost conformal overgrowth of the remaining structures takes place, that does not significantly modify further the buried QDs. This will be discussed in detail in a forthcoming publication. Fig. 3A shows Ge domes after coverage with 1 ML of Si: the comparison with the free-standing domes in Fig. 1A reveals an increase of the shallow facets (both at the top and at the base of the islands) and a corresponding shrinkage of the steeper ones. This effect is quantitatively seen in the related FP of Fig. 3B where the central {105} spots become more intense than the outer {113} and {15 3 23} ones (compare with Fig. 1C). An almost identical evolution is observed during the capping of InAs domes: after the deposition of 1ML of GaAs, four large {137} facets dominate the island shape, while the steeper {101} and {111} facets become drastically smaller (see Figs. 3D and 3E). These island shapes, that we call transition domes, represent the first step of a full backward evolution leading from domes through pyramids to shallow stepped mounds. The pristine pyramid islands follow a similar reverse evolution route but, being smaller, flatten out and finally disappear within the first overgrown monolayers.

The shape transformations during overgrowth can be rationalized by considering that at the experimental temperatures a segregation of Ge (InAs) takes place during Si (GaAs) capping because of the lower surface energy of the former material with respect to the latter (*29, 30*). As a consequence, even if bulk diffusion is thermally almost inhibited, incorporation of Si (GaAs) into the dots occurs close to the surface, resulting in a local decrease of the elastic strain energy. The net result for the islands is a tendency of reducing the extension of steep facets (that are more effective in releasing strain) in favor of shallow ones, which is efficiently done by transferring material from the top to the base of the island. This is the origin of the shapes in Figs. 3A and 3C and of the simultaneous reduction of the island heights. By continuing capping, the total strain of the islands further decreases, favoring first a pyramid and then a (001) stepped mound shape, as experimentally observed (*24*). In passing we remark that the 1 ML overgrown Ge domes in Fig. 3A are almost identical to the intermediate structures that can be observed in the pyramid-to-dome transition during growth (*22*). Therefore, the analogy between the two material systems implies that the corresponding InAs transition domes should be similar to the islands in Fig. 3C.

For the two main representative systems in semiconductor lattice-mismatched heteroepitaxy we have shown that the spontaneously forming 3D islands present only two well-defined shapes, pyramids and domes. These results agree with the theoretical model of Daruka et al. (*19*) which indeed predicts universal faceted island shapes, independently of specific material parameters. Also the overgrowth scenario that emerges from our measurements, in which the QD capping can be described as a backward transition from steeper domes to shallower pyramids, is material-independent as long as the surface energy of the cap material is higher than that of the island material. Our measurements therefore suggest that the unified picture we have presented here for the prototype systems Ge/Si(001) and InAs/GaAs(001), extends, at least qualitatively, to a large number of material combinations that follow the Stranski-Krastanow growth mode. This universal description of the growth and overgrowth processes of self-organized semiconductor quantum dots will be a valuable tool in the design and engineering of QD structures.

**Figure captions**

**Fig. 1.**

Self-organized pyramid and dome islands for different material systems. **(A)** STM topography of 7.0 ML Ge deposited on Si(001) at 550°C with a rate of 0.3 ML/s (grayscale corresponds to the local surface slope). Pyramids and domes are marked by red and green lines, respectively. **(B), (C)** Corresponding facet plots selectively evaluated for the pyramids (B) and the domes (C) only. These intensity plots result from representing in a two-dimensional histograms the values of the local surface gradient **n** = $\nabla f$, where f(x,y) is the surface height at position (x,y). The horizontal and vertical axes correspond to $\partial f/\partial x$ and $\partial f/\partial y$, where x and y stand for the [100] and [010] direction in (B) and (C) and for [110] and [1-10] in (E) and (F), respectively. The (0,0) point at the center of the plots corresponds to the {001} orientation. **(D)** STM topography of 1.8 ML InAs deposited on GaAs(001) at 500°C with a rate of 0.008 ML/s. **(E), (F)** related facet plots for pyramid (E) and dome islands (F) respectively. The main surface orientations emerging from the facet plot analysis are explicitly indicated.

**Fig. 2.**

Structural models of pyramid and dome islands. **(A)** Ge/Si(001). **(B)** InAs/GaAs(001). The different facets are marked by different gray tones.

**Fig. 3.**

Morphology of dome islands after the overgrowth with 1 ML of cap material. **(A)** 1 ML Si on Ge domes. **(B)** Corresponding facet plot. The comparison with Fig. 1C clearly shows that the intensity of the shallow {105} facets has increased at the expense of the {113} and {15 3 23} ones. **(C)** Structural model of overgrown domes. **(D), (E), (F)** An identical evolution is observed for the InAs/GaAs(001) case: InAs domes overgrown by 1 ML GaAs (D), facet plot (E), and structural model (F).

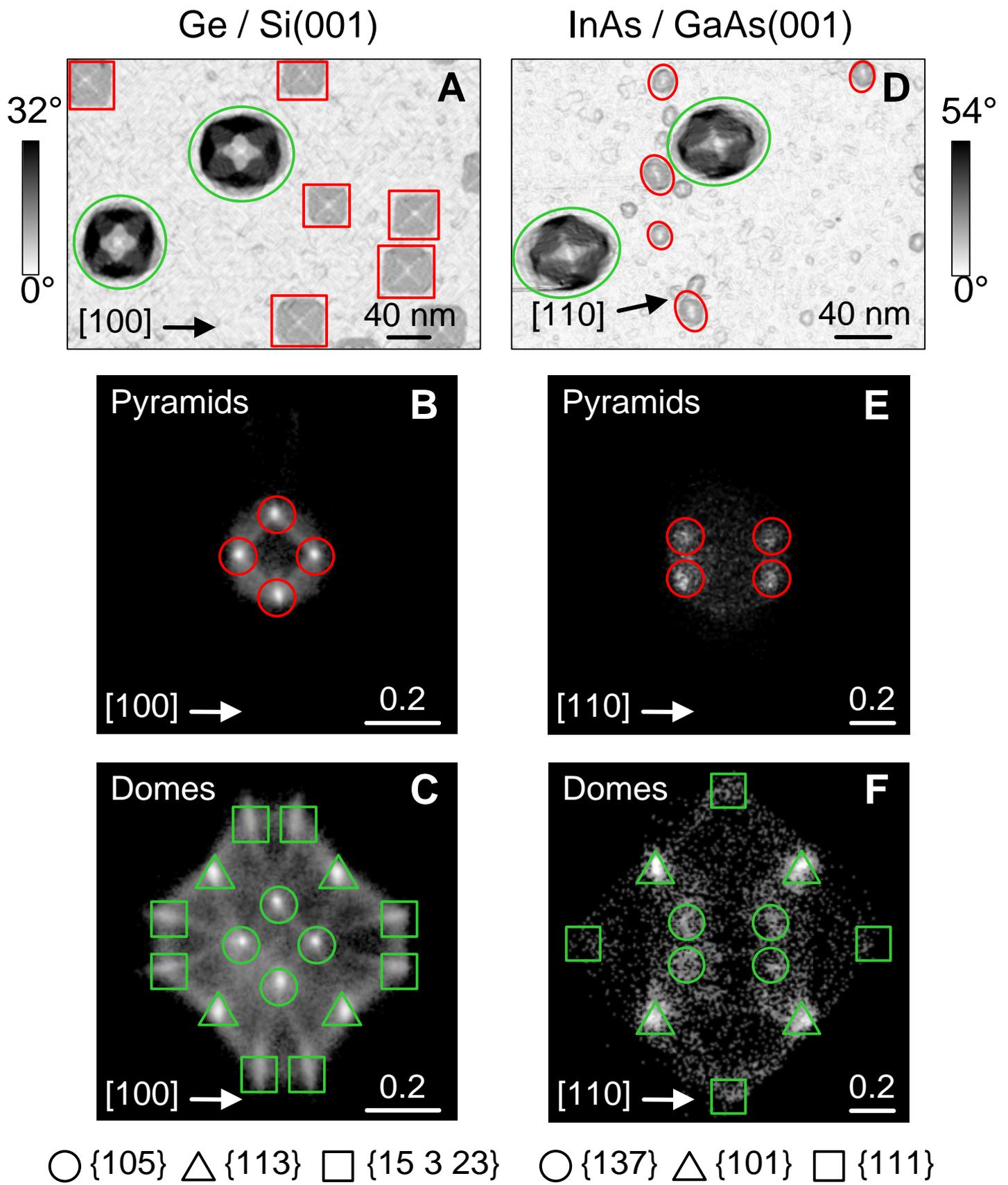

Figure 1

**A**    Ge / Si(001)

Pyramid      Dome

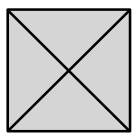 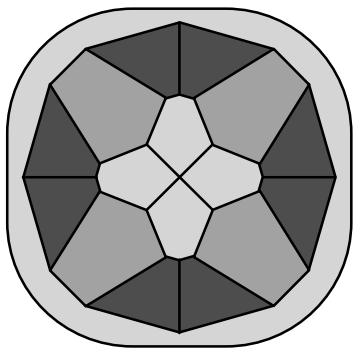

[100] ⟶

☐ {105}   ▨ {113}   ■ {15 3 23}

**B**    InAs / GaAs(001)

Pyramid      Dome

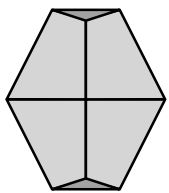 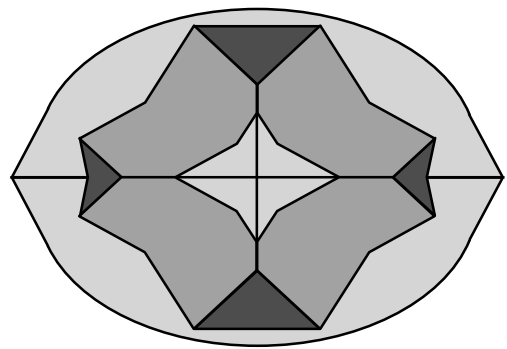

[110] ⟶

☐ {137}   ▨ {101}   ■ {111}

Figure 2

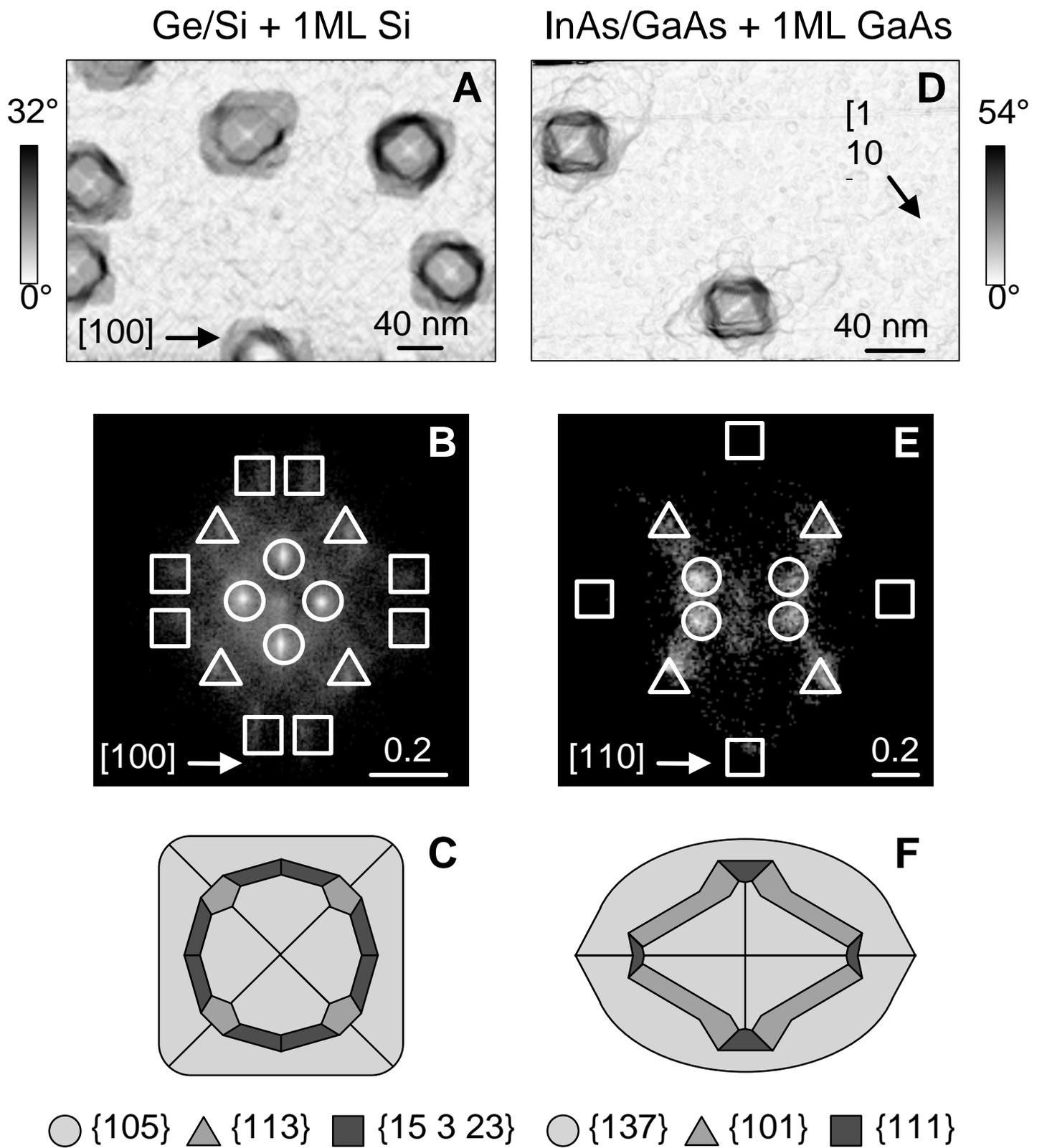

Figure 3